\documentclass[conference]{IEEEtran}
\IEEEoverridecommandlockouts
\usepackage{cite}
\usepackage{amsmath,amssymb,amsfonts}
\usepackage{algorithmic}
\usepackage{graphicx}
\usepackage{textcomp}
\usepackage{xcolor}
\usepackage{comment}
\usepackage{url}
\usepackage{hyperref}
\newcommand\tab[1][1cm]{\hspace*{#1}}

\def\BibTeX{{\rm B\kern-.05em{\sc i\kern-.025em b}\kern-.08em
    T\kern-.1667em\lower.7ex\hbox{E}\kern-.125emX}}

\begin{document}

\title{Deep Audio Analyzer: a Framework to Industrialize the Research on Audio Forensics
}

\author{
\IEEEauthorblockN{Valerio Francesco Puglisi}
\IEEEauthorblockA{
valerio.puglisi@phd.unict.it }
\and

\IEEEauthorblockN{Oliver Giudice}
\IEEEauthorblockA{
giudice@dmi.unict.it}

\and

\IEEEauthorblockN{Sebastiano Battiato}
\IEEEauthorblockA{
battiato@dmi.unict.it}
\and

\IEEEauthorblockA{
\tab[6cm]\textit{Department of Mathematics and Computer Science} \\
\tab[6cm]\textit{University of Catania}\\
\tab[6cm]\textit{Viale Andrea Doria 6, Catania, 95123, Sicily, Italy}\\
}

}

\maketitle

\begin{abstract}
Deep Audio Analyzer is an open-source speech framework that aims to simplify the research and the development process of neural speech processing pipelines, allowing users to conceive, compare and share results in a fast and reproducible way. This paper describes the core architecture designed to support several tasks of common interest in the audio forensics field, showing possibility of creating new tasks thus customizing the framework. By means of Deep Audio Analyzer, forensics examiners (i.e. from Law Enforcement Agencies) and researchers will be able to visualize audio features, easily evaluate performances on pre-trained models, to create, export and share new audio analysis workflows by combining deep neural network models with few clicks. 
One of the advantages of this tool is to speed up research and practical experimentation, in the field of audio forensics analysis thus also improving experimental reproducibility by exporting and sharing pipelines. 
All features are developed in modules accessible by the user through a Graphic User Interface. 

\end{abstract}


\section{Introduction}
Applications of novel deep learning solutions to forensics investigations has experienced unprecedented growth in interest and obtained results \cite{bib1,b1,bib2,b2,bib3,b3}, with many researchers developing innovative algorithms and models to solve complex problems. 
Research is mostly data-driven and having a lot of data gives the opportunity to find new features or more forensically "fingerprints" \cite{b22,b24,b25} which are of utter importance when dealing with the fight of Artificial Intelligence generated evidences \cite{b23}.
However, reproducing published experiments and results remain a significant challenge due to the needed programming skills required. This challenge is further compounded by the lack of (or an extremely limited) standardization in the way experiments are conducted. This issue results in a significant amount of time being spent by researchers trying to get other researchers’ code to work, which leads to a significant waste of resources. The development of speech-processing technologies has been largely driven by open-source toolkits \cite{bib4,bib5,bib6,bib7,bib8}. However, with the emergence of general-purpose deep learning libraries like TensorFlow \cite{bib9} and PyTorch \cite{bib10}, more flexible speech recognition frameworks have emerged, such as DeepSpeech \cite{bib11}, RETURNN \cite{bib12}, PyTorch-Kaldi \cite{bib13}, Espresso \cite{bib14}, Lingvo \cite{bib15}, Fairseq \cite{bib16}, ESPnet \cite{bib17}, NeMo \cite{bib18}, Asteroid \cite{bib19}, Speechbrain \cite{bib20} and hub where scientists load trained models for others to download \cite{bib21}. While it can be challenging for non-experts users to prototype new deep learning methods, as it requires knowledge of codingg and environmental setup. In this paper, we aim to highlight the importance of reproducibility in data science and discuss a solution to address the problem of knowledge of programming in different languages. Deep Audio Analyzer is a tool that enables users to visualize audio features, evaluate the performance of pre-trained models, and create new audio analysis workflows by combining deep neural network models. Through the use of Deep Audio Analyzer, users can perform these features without the need to develop any code. The tool also provides dedicated modules to test state-of-the-art models on customized data and also combine models to create a new deep learning audio processing pipeline, combing for tasks such as Automatic Speech Recognition, Speech Enhancement, Speaker Separation, Speaker Verification and Voice Activity Detection. Code is available at \href{https://github.com/valeriopuglisi/deep-audio-analyzer}{https://github.com/valeriopuglisi/deep-audio-analyzer}.

The remainder of the paper is organized as follows. 
In Section 2, we delve into the technologies comprising the architecture of the Audio Analyzer. 
Section 3 reports the proposed features and modules developed in Deep Audio Analyzer. 
Section 4 presents the considered experiments and discusses the obtained
results. Section 5 concludes the paper and proposes future works.

\section{Architecture}\label{sec3}
\begin{figure}[t!]
\centerline{\includegraphics[scale=0.4]{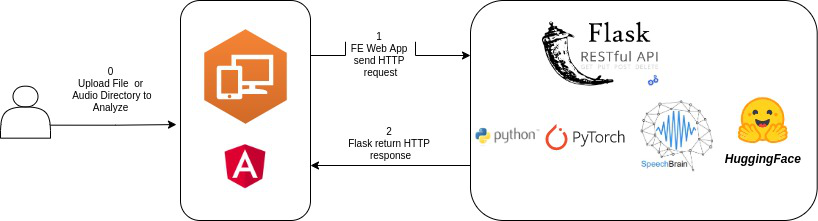}}
\caption{ \textbf{Angular Front End:} The User Interface of Deep Audio Analyzer is divided into pages and
components in order to categorize all the functions separately. All the module of Deep Audio Analyzer is developed on different pages. \textbf{Flask Backend:}
Deep Learning Audio Analyzer employs a simple software stack (i.e., Python → PyTorch → SpeechBrain → HuggingFace → Flask → Angular) to avoid dealing with too many levels of abstraction. It is developed on top of SpeechBrain and HuggingFace directly, with external APIs that can retrieve the newest model uploaded from the SpeechBrain community and other Companies.}
\label{fig:Architecture}
\end{figure}

The Architecture of Deep Learning Audio Analyzer is actually composed of a Backend service where all the artificial intelligence tasks are implemented and Front-End is written in Angular and is concerned with making Audio Analysis as easy as possible as shown in Fig.\ref{fig:Architecture}.

Angular FrontEnd framework is concerned to make easier the development and maintenance of the platform while the Backend Flask RESTful API was chosen because it is fast to develop and is written in Python, which comprises the Artificial Intelligence libraries used to develop Deep Audio Analyzer platform. 


\section{Deep Audio Analyzer}

\subsection{Audio Features Visualization Module}
\begin{figure}[t!]
\centerline{\includegraphics[scale=0.4]{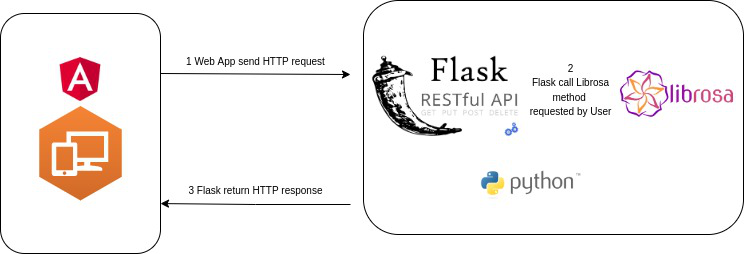}}
\caption{Architecture of Audio Feature Visualization Module.}
\label{fig2:VisualizationModule}
\end{figure}

Through the preprocessing module of Deep Audio Analyzer, it is possible to graphically analyze all the features extracted through the application of the functions present in the librosa library \cite{bib21p3} whose functions have been implemented in the Backend of the application (Fig.\ref{fig2:VisualizationModule}). 

The developed functions are shown in the next list :
\subsection{Preprocessing Audio Features}

\begin{itemize}

\item  \textbf{Linear-frequency power spectrogram:} 
The linear-frequency power spectrogram is an important tool in the field of audio forensics. It represents time on the X-axis, frequency in Hz on a linear scale on the Y-axis, and power in dB \cite{bib21p0}. It is used to identify specific events in an audio recording by allowing experts to analyze the spectral characteristics. It also enables voice analysis by studying features such as pitch, formants, and harmonics, which can aid in speaker identification and voice comparison. Furthermore, the spectrogram can reveal hidden artefacts, noise, or disturbances in an audio recording, which can then be mitigated by applying appropriate filtering techniques, thereby enhancing the desired audio content. In audio authentication, the spectrogram can be instrumental in detecting signs of audio tampering or manipulation.

\item \textbf{Log-frequency power spectrogram:} Such features can be obtained from a spectrogram by converting the linear frequency axis (measured in Hertz) into a logarithmic axis (measured in pitches).
Its logarithmic representation of frequency content enables experts to extract unique voice characteristics, classify sounds, segment audio recordings, enhance transcription accuracy, and detect potential tampering or manipulation.
The resulting representation is also called log-frequency spectrogram. 

\item \textbf{Chroma STFT:} 
Chroma STFT features are useful for analyzing the harmonic content of an audio signal and can be used in a variety of applications such as music information retrieval, audio classification, and speech recognition. They provide a way to represent the pitch content of an audio signal in a compact and efficient way and can be used to compare and classify different audio signals based on their harmonic content. 
Chroma STFT is a useful tool in audio signal processing for analyzing the chromatic content of an audio signal and can be used in a wide range of applications. This implementation is derived from chromagram E \cite{bib21p1}

\item \textbf{Chroma CQT :} The Constant-Q chromagram is a type of chroma feature representation commonly used in music analysis and processing. It is based on the Constant-Q transform, which is a frequency-domain transformation that uses a logarithmic frequency scale that approximates the way that humans perceive sound. \cite{bib21p01}

\item \textbf{Chroma CENS:} Computes the chroma variant “Chroma Energy Normalized” (CENS)\cite{bib21p2}.

 CENS features are robust to dynamics, timbre and articulation, thus these are commonly used in audio matching and retrieval applications .
 
 \item \textbf{Melspectrogram:} Compute a mel-scaled spectrogram. If a spectrogram input S is provided, then it is mapped directly onto the mel basis by $mel_f.dot(S)$. If a time-series input y, sr is provided, then its magnitude spectrogram S is first computed, and then mapped onto the mel scale by $mel_f.dot(S**power)$. By default, power=2 operates on a power spectrum. \cite{bib21p3}

\item \textbf{Mel-frequency spectrogram}: 
Display of mel-frequency spectrogram coefficients, with custom arguments for mel filterbank construction (default is $f_{max}=sr/2$).
Mel-frequency spectrograms are valuable in forensic audio analysis for visualizing and analyzing the characteristics of an audio recording relevant to a legal case. They help identify specific sounds, voices, recording quality, and potential tampering. By extracting features like pitch, spectral content, and temporal characteristics, it enables comparisons between different recordings to determine their common source. 
They are useful for speaker identification, voice matching, and background noise analysis. 
Mel-frequency spectrograms provide a perceptually relevant representation of audio and allow forensic analysts to determine important details about the origin and authenticity of recordings.

\item \textbf{Mel-frequency cepstral coefficients (MFCCs)}: Mel-frequency cepstral coefficients (MFCCs) are a type of feature representation commonly used in audio signal processing and analysis, particularly in speech recognition and forensic audio analysis. MFCCs are derived from the Mel-frequency spectrogram, which is a spectrogram that uses a frequency scale that is more aligned with human perception of sound. In forensic audio analysis, MFCCs can be used as a feature representation to compare and analyze different audio recordings. 
By computing the MFCCs for different segments of an audio recording, forensic audio analysts can identify characteristic patterns and features that may be relevant to a legal case. 

\item \textbf{Compare different DCT bases}:	In audio signal processing, the discrete cosine transform (DCT) is a widely used method for transforming time-domain audio signals into a frequency-domain representation. 
There are different types of DCTs that use different basis functions, or sets of orthogonal functions, to represent the signal in the frequency domain. 
The choice of DCT basis functions depends on the specific application and the trade-offs between computational efficiency, frequency resolution, and energy compaction. In many cases, the standard DCT-II is a good choice for audio signal processing applications, but other DCT bases may be more appropriate for certain types of signals or processing tasks.

\item \textbf{Root-Mean-Square (RMS)}: Compute root-mean-square (RMS) value for each frame, either from the audio samples y or from a spectrogram S.Computing the RMS value from audio samples is faster as it doesn’t require an STFT calculation. However, using a spectrogram will give a more accurate representation of energy over time because its frames can be windowed, thus prefer using S if it’s already available.	

\item \textbf{Spectral Centroid}: Compute the spectral centroid.Each frame of a magnitude spectrogram is normalized and treated as a distribution over frequency bins, from which the mean (centroid) is extracted per frame \cite{bib21p4}. 
The spectral centroid is a measure used in audio forensics to characterize an audio signal, often indicating the perceived "brightness" of a sound. It aids in differentiating sounds and identifying unique voices, thus assisting in speaker identification. The spectral centroid can also reveal potential audio tampering, as inconsistencies might suggest alterations. 

\item \textbf{Spectral Bandwidth}: Compute $p’th-order$ spectral bandwidth \cite{bib21p4}. 
In the realm of audio enhancement, knowledge of the spectral bandwidth can aid in developing strategies to filter out unwanted components from a recording. 
It also can be used as a fingerprint, the spectral bandwidth of an individual's voice can be unique. 
This characteristic can be analyzed to potentially match a voice to a specific person, which can prove extremely useful in forensic investigations.

\item \textbf{Spectral Contrast}: Compute spectral contrast. Each frame of a spectrogram S is divided into sub-bands. For each sub-band, the energy contrast is estimated by comparing the mean energy in the top quantile (peak energy) to that of the bottom quantile (valley energy)\cite{bib21p5}. High contrast values generally correspond to clear, narrow-band signals, while low contrast values correspond to broad-band noise.

\end{itemize}

\subsection{Deep Learning Audio Inference Module}
Deep Audio Analyzer puts implements several audio analysis tasks using deep learning methods.
The neural networks present in Deep Audio Analyzer are state of the art for the different tasks, and their implementation of them is currently supported by the SpeechBrain \cite{bib20} framework that implements interfaces through which it is possible to download and execute neural network models through the HuggingFace \cite{bib21} aggregator.
Table \ref{tab:1} summarizes the various neural network models for the different tasks and related datasets on which they were trained and the obtained performance.

\begin{table}[t!]
\caption{Deep Learning Models: ASR: Automatic Speech Recognition, 
ER: Emotion Recognition, 
LI: Language Identification,
SE: Speech Enhancement, 
SS: Speech Separation, 
SV: Speaker Verification, 
VAD: Voice Activity Detection  }  
\begin{tabular}{|l|l|l|l|}
\hline
\textbf{Task} & 
\textbf{System} & 
\textbf{Dataset} &
\textbf{Performance}\\
\hline 
ASR &
\begin{tabular}[l]{@{}l@{}} wav2vec2\cite{bib22} \end{tabular} & 
\textit{LibriSpeech}\cite{bib23}& 
WER=1.90\%\\ 
\hline 
ASR &
\begin{tabular}[l]{@{}l@{}l@{}} CNN + \\ Transformer \end{tabular} & 
\textit{LibriSpeech}\cite{bib23}& 
WER=2.46\% \\ 
\hline 
ASR &
\begin{tabular}[l]{@{}l@{}l@{}} CRDNN + \\ distillation \end{tabular} & 
\textit{TIMIT}\cite{bib24}& 
PER=13.1\%\\ 
\hline 
ASR &
\begin{tabular}[l]{@{}l@{}l@{}} CRDNN + \\ RNN+ LM \end{tabular} & 
\textit{Librispeech}\cite{bib24}& 
\begin{tabular}[l]{@{}l@{}l@{}} WER=3.09\% \\(test-clean)
\end{tabular}\\
\hline 
ASR &
\begin{tabular}[l]{@{}l@{}l@{}} Conformer + \\ Transf. LM \end{tabular} & 
\textit{Librispeech}\cite{bib24}& 
\begin{tabular}[l]{@{}l@{}l@{}} WER=3.09\% \\(test-clean)
\end{tabular}\\
\hline 
ASR &
\begin{tabular}[l]{@{}l@{}l@{}} CRDNN + \\ Transf. LM \end{tabular} & 
\textit{Librispeech}\cite{bib24}& 
\begin{tabular}[l]{@{}l@{}l@{}} WER=8.51\% \\(test-clean)
\end{tabular}\\
\hline 
ASR &
\begin{tabular}[l]{@{}l@{}l@{}} wav2vec2 + \\ CTC/Att.\cite{bib22, bib25} \end{tabular} & 
\textit{TIMIT}\cite{bib24}&
PER=8.04\%\\ 

\hline 
ASR &
\begin{tabular}[l]{@{}l@{}l@{}} wav2vec2 + \\ CTC \end{tabular} & 
\textit{CV (English)}\cite{bib26}&
WER=15.6\% \\ 
\hline 
ASR &
\begin{tabular}[l]{@{}l@{}l@{}} wav2vec2 + \\ CTC \end{tabular} & 
CV (German)\cite{bib26}           &
WER=9.54\%    \\
\hline 
ASR &
\begin{tabular}[l]{@{}l@{}l@{}} wav2vec2 + \\ CTC \end{tabular} & 
CV (French)\cite{bib26}           &
WER=9.96\%    \\
\hline 
ASR &
\begin{tabular}[l]{@{}l@{}l@{}} wav2vec2 + \\ seq2seq \end{tabular} & 
CV (Italian)\cite{bib26}          & 
WER=9.86\%    \\ 
\hline 
ASR &
\begin{tabular}[l]{@{}l@{}l@{}} wav2vec2 + \\ seq2seq \end{tabular} & 
AISHELL\cite{bib27}             &
CER=5.58\%    \\
\hline 
ER &  
wav2vec                      & 
IEMOCAP\cite{bib35}                     & 
Acc.=79.8\% \\ 
\hline
ER &
wav2vec                      & 
CommonLang.\cite{bib35}                     & 
Acc.=84.9\% \\ 
\hline
LI &
ECAPA-TDNN\cite{bib33} &
CommonLang.\cite{bib36} &
Acc.=84.9\%\\ 
\hline
SE &
MetricGAN+&
VoiceBank &
\begin{tabular}[l]{@{}l@{}}PESQ=3.08 (test)\end{tabular} \\ 
\hline
SE&
SepFormer&
WHAMR!&
\begin{tabular}[l]{@{}l@{}l@{}}SI-SNR= 10.59, \\  PESQ=2.84 (test)\end{tabular} \\ 
\hline
SE&
SepFormer&
WHAM! (8k)&
\begin{tabular}[l]{@{}l@{}l@{}}SI-SNR= 14.35, \\ PESQ=3.07 (test)\end{tabular} \\ 
\hline
SE&
SepFormer&
WHAM! (16k)&
\begin{tabular}[l]{@{}l@{}l@{}}SI-SNRi 13.5 dB, \\ SDRi= 13.0 dB\end{tabular} \\ 
\hline
SS &
SepFormer\cite{bib28} &
WSJ2MIX\cite{bib29} & 
SDRi=22.6 dB  \\ 
\hline 
SS &
SepFormer\cite{bib28} & 
WSJ3MIX\cite{bib29} & 
SDRi=20.0 dB  \\ 
\hline 
SS &
SepFormer\cite{bib28} & 
WHAM!\cite{bib30} & 
SDRi= 16.4 dB \\ 
\hline 
SS &
SepFormer\cite{bib28} & 
WHAMR!\cite{bib31} & 
SDRi= 14.0 dB \\ 
\hline 
SS &
SepFormer\cite{bib28} &
Libri2Mix\cite{bib32} & 
SDRi= 20.6 dB  \\ 
\hline 
SS &
SepFormer\cite{bib28} &
Libri3Mix\cite{bib32} & 
SDRi= 18.7 dB\\ 
\hline
SV &
ECAPA-TDNN\cite{bib33} &
VoxCeleb2                      &
EER=0.69\%  \\ 
\hline
VAD &
CRDNN\cite{bib33} &
LibriParty        &
F-score=0.94\%  \\ 
\hline
\multicolumn{4}{l}{
Deep Learning Audio Analysis Features.
\label{tab:1}.}
\end{tabular}
\end{table}

\subsection{Pipeline Creation and Saving}
\begin{figure}[t!]
        \centerline{\includegraphics[scale=0.42]{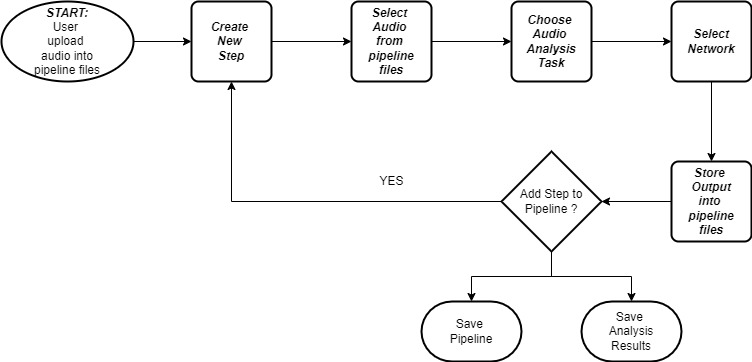}}
	\caption{Pipeline Creation and Dynamic Audio Analysis Flowchart}
	\label{fig:DLAA_architecture_pipeline_creation}
\end{figure}

Through Deep Audio Analyzer, it is possible to perform analysis of an audio file dynamically by creating an audio analysis pipeline. Fig \ref{fig:DLAA_architecture_pipeline_creation} shows the flowchart expressing the working principle of audio analysis with Deep Audio Analyzer.

The following list represents the process of analysis and pipeline creation:
\begin{enumerate}
    \item First, the input audio file is selected, 

    \item Once the file is selected, the task to be performed and consequently, the neural network model is chosen from those available for that task, 

    \item Once the step is defined, it is possible to execute it by means of a POST request sent to the server, which will execute the neural network in inference and return the result of the task performed to the client 


    \item Then it is possible to add a new step to the pipeline by choosing on which file to perform the analysis or save the pipeline from executing it later on different files
    
\end{enumerate}

\subsection{Pipeline Execution and Download Report}
\begin{figure}[t!]
        \centerline{\includegraphics[scale=0.42]{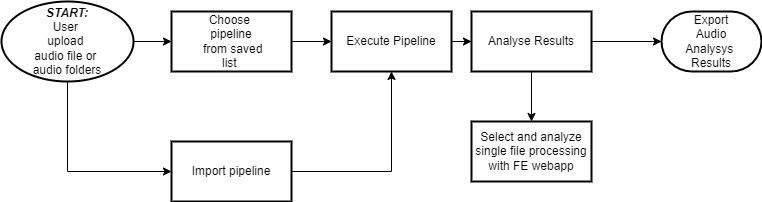}}
	\caption{Pipeline Execution Audio Analysis Flowchart}
	\label{fig:DLAA_architecture_pipeline_execution}
\end{figure}

Audio analysis pipelines that have been previously saved by the user are available in the ”pipelines” section. In this section, it is possible to run a previously created pipeline, on one or more (previously recorded) audio files, or to perform a recording of an audio file using the application GUI. Once the type of input to be analyzed has been selected, it is possible to choose the type of pipeline and view its steps. Deep Audio Analyzer will then display
via Frontend the results of the inferences performed on the Backend side as described in the previous paragraphs. After the analysis process, it is possible to download the reports containing the pipeline executed on each file and its results for each step that is part of
it. Figure \ref{fig:DLAA_architecture_pipeline_execution} describes the flowchart for pipeline execution and reporting.

\section{Experiments and Results}
In this section, some examples of new pipelines and the tests performed on the performance of the neural networks available for the different tasks and the obtained results using the Deep Audio Analyzer, are described.

\subsection{Example of new pipeline}
In this section, we present two examples of pipeline creation that can be used for investigative purposes in interception contexts. The first example concerns the transcription of speech from multiple people speaking different languages, while the second example concerns the transcription of speech in noisy environments using speech enhancement models.

\subsubsection{Multi-speaker Multi-Language ASR : Speech Separation + Language ID + ASR}
This pipeline dedicated to transcribing speech in different languages from a maximum of three speakers is composed of the following steps:
\begin{enumerate}
    \item Addition of the file of interest, selection of the audio separation model, and execution.
    \item This step provides three output files; thus, it is necessary to create three new voice activity detection (VAD) steps one per output file of step one.
    \item The three steps will each output a file where silence has been removed. It will then be necessary to introduce three additional steps with our implemented language identification and automatic speech recognition module, taking as input the audio processed with VAD in the previous steps.
\end{enumerate}

\subsubsection{Automatic Speech Recognition in noisy environment}

In forensic investigations, it is often necessary to transcribe highly noisy audio. This context is often overlooked in academic settings, as the focus is on evaluating transcriptions in clean or low-noise/echo environments. Therefore, creating a pipeline that involves the use of enhancement models and voice activity detection improves the results of automatic speech recognition. The creation of this pipeline consists of the following steps :
\begin{enumerate}
    \item Loading the audio file and selecting the desired model for the speech enhancement task. 
    \item Adding a new step that takes as input the improved signal produced by step 1 and select the desired Voice Activity Detection model. 
    \item Adding the language identification and automatic speech recognition task, implemented by us.
\end{enumerate}

\subsection{Experiments}
Deep Audio Analyzer is an application designed as a support tool for audio analysis in forensic and also academic fields.
For these reasons, several experiments have been implemented including validating models related to different tasks on different datasets through the implementation of appropriate evaluation metrics using the library \cite{bib37}.
\\
\subsubsection{Validate performance of a pre-trained neural network on different task}
The first test case consists of evaluating by means of the metrics set out in the introduction chapter, the behaviour of the various networks with datasets that are different from the training datasets, but which have been realised for the same task to see how are robust the networks, varying the datasets for the same type of task.
For example, to evaluate the performance of the Automatic Speech Recognition networks in Deep Audio Analyzer. The current Automatic Speech Recognition networks are trained mainly on Librispeech, Voxpopuli and Common Voice and it is possible to see the performance on different datasets in the table \ref{tab:EvaluationASR} by the implementation of Character Error Rate (CER) and Word Error Rate (WER) \cite{bib38, bib39}.
We also implemented evaluation for speech separation models in table \ref{tab:EvaluationSpeechSeparation} by the implementation of five differents metrics: Source .

\subsubsection{User validation of the performance of the deep neural networks available for a given task }
Suppose we wanted to test the quality of the neural networks available in the automatic speech recognition application for a specific language, using files not belonging to datasets. 
It is possible to do this by the pipeline creation section.
The user creates a pipeline and add as many steps as necessary to compare the neural networks available for that language and save the pipeline.
Then the users can run the various comparison pipelines (previously created) to test the behaviour of the various networks for tasks in examples that are not included in the training datasets.
In this use case, it is not possible to perform a validation according to the metrics related to the task being analysed, because the relevant ground-thoughts are missing. 
For this reason, it was decided to predefine a perceptual quality index ranging from 1 to 10 for the tasks on the platform.

Subsequently, a pipeline was created for each task in order to compare all the available networks in a single process and then manually evaluate the performance of the individual network from 1 to 10.
In this way, it is easy to sample perceptual opinions from experts in the field in order to assess robustness not only in the various existing datasets for the generic task but also with audio files recorded in real 'into the wild' situations.

\subsection{Results}
\subsubsection{Model Evaluation on different Datasets}
Tables \ref{tab:EvaluationASR}, \ref{tab:EvaluationSpeechSeparation} show the Evaluation module applied Automatic Speech recognition task and Speech Separation task with pre-trained models on some datasets. 
However, evaluations conducted on different datasets show that even though a network may show good performance on the training dataset, it may not perform well on other data from different contexts.
With Deep Audio Analyzer is possible to upload customized trained models in order to achieve better performance on private datasets.
\begin{table}[t!]
\caption{Evaluation of Automatic Speech Recognition Models on different Test Datasets}  
\begin{tabular}{|l|l|l|}
\hline
\textit{\textbf{System}} &
\textit{\textbf{\begin{tabular}[l]{l@{}l@{}}Training dataset \&\\ Evaluation Metrics\end{tabular}}} &
\textbf{\begin{tabular}[l]{l@{}l@{}}Test dataset \&\\ Evaluation Metrics\end{tabular}} 
\\ 
\hline 
\begin{tabular}[l]{l@{}l@{}}wa2vec2 +\\ CTC \end{tabular} & 
\begin{tabular}[l]{l@{}l@{}}Voxpopuli DE\\ WER=18.91\%\end{tabular} & 
\begin{tabular}[l]{l@{}l@{}l@{}}CommonVoice 10 DE \\ CER=20.13\% \\ WER=53.15\%\end{tabular} 
\\ 
\hline
\begin{tabular}[l]{l@{}l@{}}CRDNN with \\  CTC/Attention \end{tabular} & 
\begin{tabular}[l]{l@{}l@{}}Voxpopuli DE\\ WER=9.86\%\end{tabular} &
\begin{tabular}[l]{l@{}l@{}l@{}}CommonVoice 10 DE \\ CER=?\% \\ WER=?\%\end{tabular}
\\ 
\hline
\begin{tabular}[l]{l@{}l@{}}CRDNN with \\ Transformer LM \end{tabular} & 
\begin{tabular}[l]{l@{}l@{}}Librispeech EN \\ WER=8.51\%\end{tabular} & 
\begin{tabular}[l]{l@{}l@{}l@{}}CommonVoice 10 EN \\ CER=25.08\% \\WER=47.37\%\end{tabular} 
\\ 
\hline
\begin{tabular}[l]{l@{}l@{}l@{}}CRDNN +\\ RNN+ \\ LM \end{tabular} & 
\begin{tabular}[l]{l@{}l@{}l@{}}Librispeech EN \\ CER=??? \\ WER=??? \end{tabular} & 
\begin{tabular}[l]{l@{}l@{}l@{}}CommonVoice 10 EN \\ CER=28.88\% \\ WER=50.05\% \end{tabular} 
\\ 
\hline
\begin{tabular}[l]{l@{}l@{}}wa2vec2 +\\ CTC \end{tabular} & 
\begin{tabular}[l]{l@{}l@{}l@{}}Librispeech EN \\ CER=??? \\ WER=15.69\% \end{tabular} & 
\begin{tabular}[l]{l@{}l@{}l@{}}CommonVoice 10 EN \\ CER= 19.78\% \\ WER=32.06\% \end{tabular} 
\\ 
\hline
\begin{tabular}[l]{l@{}l@{}}wa2vec2 +\\ CTC \end{tabular} & 
\begin{tabular}[l]{l@{}l@{}l@{}}Vocpopuli EN \\ CER=???\% \\ WER=???\% \end{tabular} & 
\begin{tabular}[l]{l@{}l@{}l@{}}CommonVoice 10 EN \\ CER=32.03\% \\ WER= 64.52\% \end{tabular} 
\\ 
\hline
\begin{tabular}[l]{l@{}l@{}}wa2vec2 +\\ CTC \end{tabular} & 
\begin{tabular}[l]{l@{}l@{}l@{}}Vocpopuli ES \\ CER=???\% \\ WER=15.69\% \end{tabular} & 
\begin{tabular}[l]{l@{}l@{}l@{}}CommonVoice 10 ES \\ CER=17.2822\% \\ WER=46.31\% \end{tabular} 
\\ 
\hline
\begin{tabular}[l]{l@{}l@{}}wa2vec2 +\\ CTC \end{tabular} & 
\begin{tabular}[l]{l@{}l@{}l@{}}Vocpopuli FR \\ Test CER=3.19 \\ WER=9.96\% \end{tabular} & 
\begin{tabular}[l]{l@{}l@{}l@{}}CommonVoice 10 FR \\ CER=25.97\% \\ WER=58.70\% \end{tabular} 
\\ 
\hline
\begin{tabular}[l]{l@{}l@{}}CRDNN with \\  CTC/Attention \end{tabular} & 
\begin{tabular}[l]{l@{}l@{}l@{}}CommonVoice 9 FR\\ CER=6.54\%, \\ WER=17.70\%\end{tabular} &
\begin{tabular}[l]{l@{}l@{}l@{}}CommonVoice 10 FR\\ CER= 9.55\% \\ WER=30.82\% \end{tabular}
\\ 
\hline
\begin{tabular}[l]{l@{}l@{}}CRDNN with \\  CTC/Attention \end{tabular} & 
\begin{tabular}[l]{l@{}l@{}l@{}}CommonVoice 9 IT\\ CER=5.40\% \\ WER=16.61\% \end{tabular} &
\begin{tabular}[l]{l@{}l@{}l@{}}CommonVoice 10 IT\\ CER=7.78\% \\ WER=27.69\% \end{tabular}
\\ 
\hline
\begin{tabular}[l]{l@{}l@{}}wa2vec2 \\  \end{tabular} & 
\begin{tabular}[l]{l@{}l@{}l@{}}CommonVoice 9 IT\\ Test CER=???\% \\ WER=9.86\% \end{tabular} & 
\begin{tabular}[l]{l@{}l@{}l@{}}CommonVoice 10 IT \\ CER=7.30\% \\ WER=21.66\%  \end{tabular} 
\\ 
\hline
\begin{tabular}[l]{l@{}l@{}}wa2vec2 \\  \end{tabular} & 
\begin{tabular}[l]{l@{}l@{}l@{}}VoxPopuli IT\\ Test CER=???\% \\ WER= 45.2\% \end{tabular} & 
\begin{tabular}[l]{l@{}l@{}l@{}}CommonVoice 10 IT \\ CER=16.00\% \\ WER=52.57\%  \end{tabular} 
\\ 
\hline
\end{tabular}
\label{tab:EvaluationASR}
\end{table}

\begin{table}[t!]
\caption{Evaluation of Speech Separation Models on different Test Datasets}  
\begin{tabular}{|l|l|l|}
\hline
\textit{\textbf{System}} &
\textit{\textbf{\begin{tabular}[l]{l@{}l@{}}Training dataset \&\\ Evaluation Metrics\end{tabular}}} &
\textbf{\begin{tabular}[l]{l@{}l@{}}Test dataset \&\\ Evaluation Metrics\end{tabular}} 
\\ 
\hline
\begin{tabular}[l]{l@{}} Sepformer \end{tabular} & 
\begin{tabular}[l]{l@{}l@{}}WSJ2MIX \\ SDRi=22.6 dB (test) \end{tabular} & 
\begin{tabular}[l]{l@{}l@{}l@{}l@{}l@{}l@{}l@{}} Libri2Mix 16K Min \\ SNR= -9.3865, \\ SDR = -0.2170, \\ SI-SNR = -2.5669, \\ SI-SDR= -2.5678, \\ PESQ= 2.0454, \\ STOI= 0.5051 \end{tabular} 
\\ 
\hline
\begin{tabular}[l]{l@{}} Sepformer \end{tabular} & 
\begin{tabular}[l]{l@{}l@{}}WSJ2MIX \\ SDRi=22.6 dB (test) \end{tabular} & 
\begin{tabular}[l]{l@{}l@{}l@{}l@{}l@{}l@{}l@{}} Libri2Mix 16K Max \\ SNR = -9.1042, \\ SDR = -0.0988, \\ SI-SNR = -2.0402, \\ SI-SDR = -2.0445, \\ PESQ = 2.0879, \\ STOI = 0.5297 \end{tabular} 
\\ 
\hline
\begin{tabular}[l]{l@{}} Sepformer \end{tabular} & 
\begin{tabular}[l]{l@{}l@{}}WSJ3MIX \\ SDRi=20.0 dB (test) \end{tabular} & 
\begin{tabular}[l]{l@{}l@{}l@{}l@{}l@{}l@{}l@{}} Libri3Mix 16K Min \\ SNR = -8.2628\\ SDR = -5.3410, \\ SI-SNR = -4.8382, \\ SI-SDR=  -4.8382, \\ PESQ = 1.5473, \\ STOI = 0.3136 \end{tabular} 
\\ 
\hline
\begin{tabular}[l]{l@{}} Sepformer \end{tabular} & 
\begin{tabular}[l]{l@{}l@{}}WSJ3MIX \\ SDRi=20.0 dB (test) \end{tabular} & 
\begin{tabular}[l]{l@{}l@{}l@{}l@{}l@{}l@{}l@{}} Libri3Mix 16K Max \\ SNR = -8.3537\\ SDR = -5.3429, \\ SI-SNR = -7.8382, \\ SI-SDR=  -7.8382, \\ PESQ = 1.6473, \\ STOI = 0.3903 \end{tabular} 
\\ 
\hline
\end{tabular}
\label{tab:EvaluationSpeechSeparation}
\end{table}

\section{Conclusion }
In this paper, we  described Deep Audio Analyzer, an audio analysis platform that aims to cover the entire audio analysis process. 
Deep Audio Analyzer is a framework that allows the comparison of state-of-the-art models for speech analysis with no lines of code. It enables researchers to reduce the benchmarking time of different models. 

\end{document}